\documentclass[usenatbib]{aa}
\pdfoutput=1  
\usepackage[]{graphicx}
\usepackage{color}
\usepackage{txfonts}
\usepackage{array}
\usepackage{hyperref}
\newcommand{\hii}{H\textsc{ii}}
\newcommand{\solmass}{M$_{\odot}$} 
\newcommand{\mewm}{$\mu$m} 

\newcommand{\asec}{$^{\prime\prime}$}
 
\newcommand{\meth}{CH$_3$OH}

\newcommand{\water}{H$_2$O}
\newcommand{\mma}{MM1\textit{a}}
\newcommand{\mmb}{MM1\textit{b}}

\begin{document}

   \title{Tightening the belt: Constraining the mass and evolution in SDC335}


   \author{A. Avison \inst{1,2}\fnmsep\thanks{\email{adam.avison@manchester.ac.uk}} 
   		\and N. Peretto\inst{3} 
		\and G.A. Fuller \inst{1,2}
		\and A. Duarte-Cabral \inst{4}
		\and A. Traficante\inst{1}
		\and J.E. Pineda\inst{5}  }

   \institute{ Jodrell Bank Centre for Astrophysics, School of Physics and Astronomy, University of Manchester, Manchester, M13 9PL, UK
          \and
          UK ALMA Regional Centre Node             
          \and
          School of Physics and Astronomy, Cardiff University, Queens Buildings, The Parade, Cardiff CF24 3AA, UK
          \and
          School of Physics and Astronomy, University of Exeter, Stocker Road, Exeter, EX4 4QL, UK
          \and
          ETH Z\"{u}rich,  Institut f\"{u}r Astronomie, Wolfgang-Pauli-Str. 27, Z\"{u}rich   }


 
  \abstract
   {}
   {Recent ALMA observations identified one of the most massive star-forming cores yet observed in the Milky Way;  SDC335-MM1, within the infrared dark cloud SDC335.579-0.292. Along with an accompanying core MM2, SDC335 appears to be in the early stages of its star formation process. In this paper we aim to constrain the properties of the stars forming within these two massive millimetre sources.}
   {Observations of SDC335 at 6, 8, 23 and 25~GHz were made with the Australia Telescope Compact Array. We report the results of these continuum measurements, which combined with archival data, allow us to build and analyse the spectral energy distributions (SEDs) of the compact sources in SDC335.}
   {Three Hyper Compact \hii\ regions within SDC335 are identified, two of which are within the MM1 core. For each HC\hii\ region, a free-free emission curve is fit to the data allowing the derivation of the sources' emission measure, ionising photon flux and electron density. Using these physical properties we assign each HC\hii\ region a zero-age main sequence spectral type, finding two protostars with characteristics of spectral type B1.5 and one with a lower limit of B1-B1.5.\\Ancillary data from infrared to mm wavelength are used to construct free-free component subtracted SEDs for the mm-cores, allowing calculation of the bolometric luminosities and revision of the previous gas mass estimates.}
   {The measured luminosities for the two mm-cores are lower than expected from accreting sources displaying characteristics of the ZAMS spectral type assigned to them. The protostars are still actively accreting, suggesting that a mechanism is limiting the accretion luminosity, we present the case for two different mechanisms capable of causing this. Finally, using the ZAMS mass values as lower limit constraints, a final stellar population for SDC335 was synthesised finding SDC335 is likely to be in the process of forming a stellar cluster comparable to the Trapezium Cluster and NGC6334 I(N).}


   \maketitle
\section{Introduction}

The massive infrared dark cloud (IRDC) SDC335.579-0.292 \citep{Peretto09} has been noted as an object of interest due to its large mass but low levels of radiative feedback for the protostars inhabiting it. The lack of the disruptive effects of radiative feedback allow SDC335 to be used to observe the initial conditions of star formation.

\begin{figure}
\centering
\includegraphics[scale=0.4]{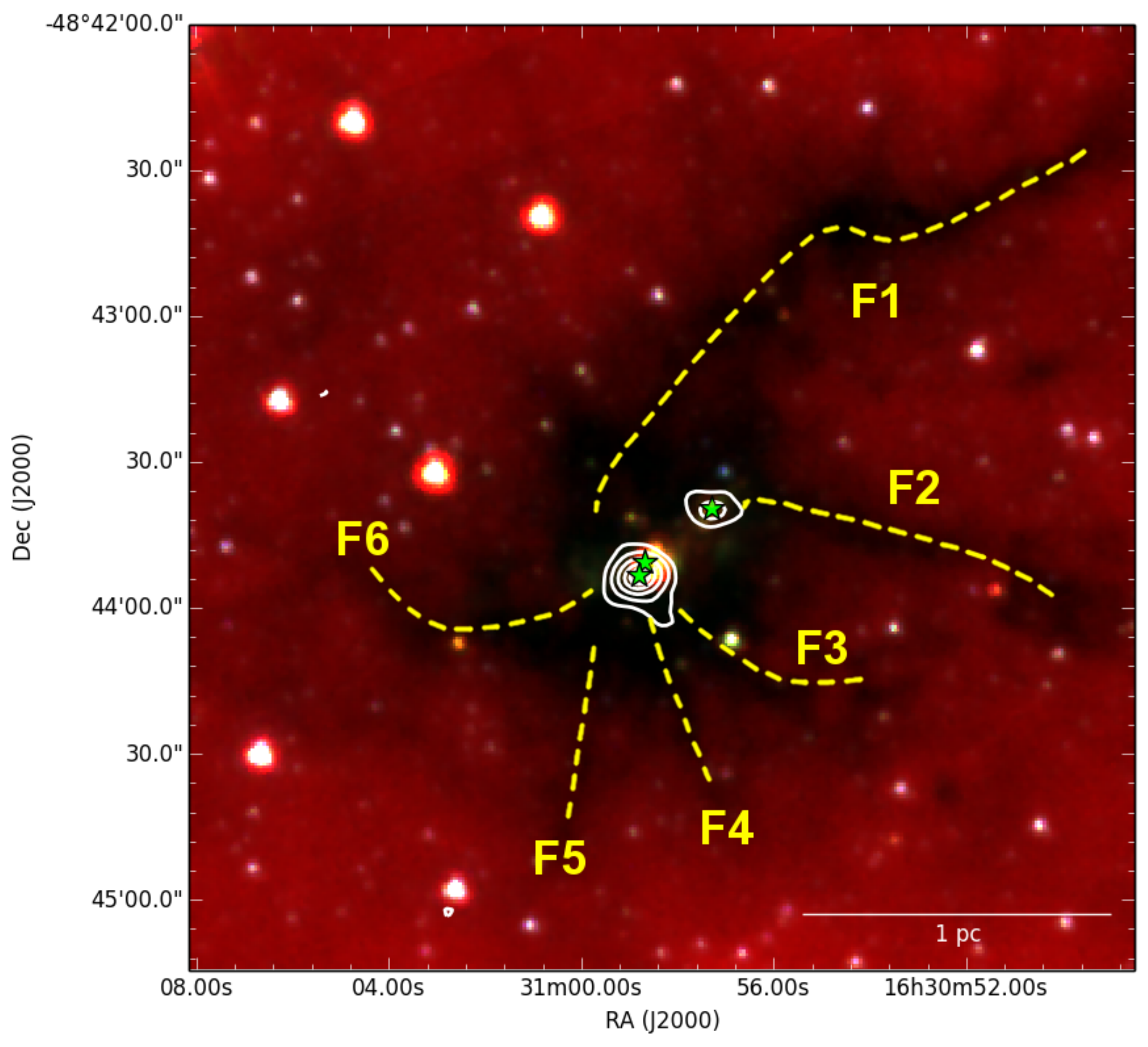}
\caption{\textit{Spitzer} three colour image of the IRDC SDC335. \textit{Colour scale}: red, green and blue GLIMPSE images at 8.0, 4.5 and 3.6\mewm\ respectively. \textit{White contours}: ALMA 3.2mm continuum, with contours at 3, 10, 25, 50, 75 and 90\% of the peak flux. \textit{Green $\star$},  Class~II  \meth\  masers. The yellow dashed lines outline the filaments of SDC335, following \citet{Peretto13}.}
\label{SDC335filaments:fig}%
\end{figure}

SDC335 shows an interesting filamentary structure (Figure \ref{SDC335filaments:fig}), with six filamentary arms radiating outward  from a denser central region. Within this central region reside two infrared bright cores, MM1 and MM2. The cores are separated by 20.2\asec\ equivalent to 0.32pc at the distance of SDC335 which throughout this work is taken as 3.25($^{+0.33}_{-0.35}$)kpc (following \citealt{Peretto13}, using the \citealt{Reid09} model). 

Both cores are associated with 6.7~GHz class II methanol masers \citep{MMB330to345}, with two methanol masers associated with MM1 and a single maser associated with MM2. The class II methanol maser is a unique tracer of massive star formation \citep{Minier03,Xu08,Breen13}, meaning that SDC335 is harbouring massive protostellar objects. However, prior to this work no free-free emission at 6cm had been detected to a sensitivity limit of  0.2mJy \citep{Garay02}, indicating that these sources are indeed young and in a pre-ionising phase of their evolution.

SDC335 was the target of ALMA Cycle 0 observations with ALMA Band 3 \citep{Peretto13}. ALMA observed the 3.2mm dust continuum as well as spectral line transitions for \meth\ (13-12) and N$_2$H$^+$ (1-0). Using these data it was shown that MM1 has a mass of $\sim$500 \solmass\ making it one of the most massive star forming cores observed in the Galaxy. MM2 was found to have a mass of $\sim$50 \solmass \footnote{We here state approximate masses for MM1 and MM2 as a calculation error has been found in \citet{Peretto13} since publication leading to $\sim$16\% smaller masses. This issue will be addressed in a forthcoming corrigendum.}. The ALMA data supplemented by \textit{Spitzer, Herschel} and \textit{Mopra} data also showed that the whole of SDC335 was undergoing global collapse, the converging network of filaments feeding the central region of the cloud with pristine gas.

In this paper we present follow up observations of SDC335 made using the Australia Telescope Compact Array in the radio (6, 8, 23 and 25~GHz) and use these new data to better constrain the masses of the stars forming in MM1 and MM2. In addition, we discuss the implications of the presence of three newly identified HC\hii\ regions associated with these millimetre sources.

\section{Observations}
\subsection{ATCA}
The primary source of data for this work is the Australia Telescope Compact Array (ATCA), from two dedicated observing runs under project code C2835. 

\begin{table}
\caption[]{Observing characteristics of interferometric studies of SDC335 used in this study.}
\begin{center}
\begin{tabular}{c r c c}
\hline
\hline
 & & Continuum & Synthesised \\
Telescope & Freq. &  Sensitivity & Beam  \\
 &  [GHz] &  [mJy bm$^{-1}$] & [\asec $\times$ \asec]\\
\hline \noalign {\smallskip}
ALMA & 93.7 & 0.400 & 5.6$\times$4.0 \\
ATCA & 25.0 & 0.023 & 1.9$\times$1.0 \\
ATCA & 23.0 & 0.026 &  2.1$\times$1.1\\
ATCA & 8.0 & 0.008 & 2.3$\times$1.0 \\
ATCA & 6.0 & 0.013 & 2.9$\times$1.4 \\
\hline
\end{tabular}
\end{center}
\label{ObsParams:tab}
\end{table}

\begin{figure}
\centering
\includegraphics[scale=0.35]{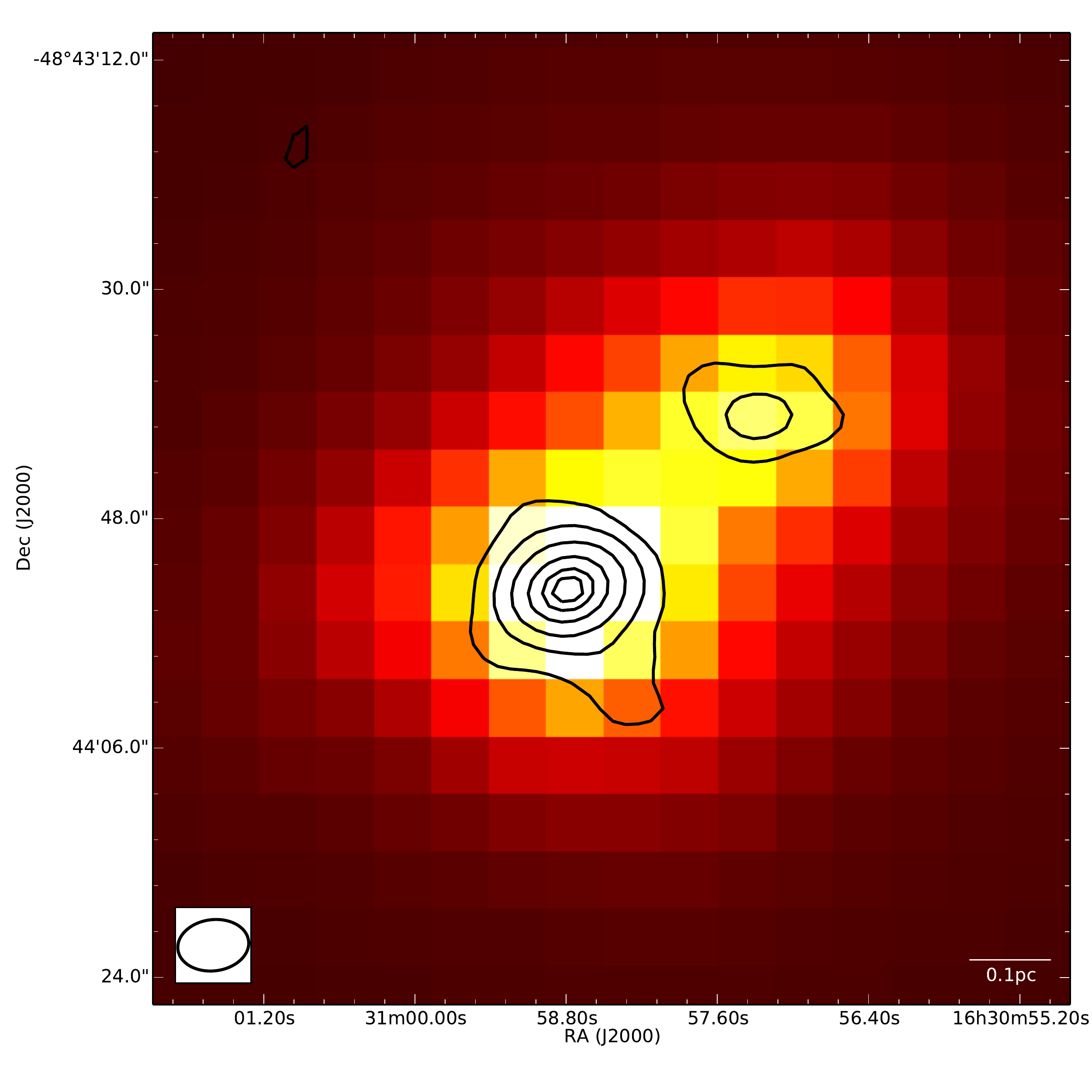}
\caption{PACS 160\mewm\ image of SDC335, overlaid with ALMA 3.2mm data (contours). It is clear that MM1 and MM2 are blended in this image with emission from MM1 possibly affecting photometry for MM2. The ALMA synthesised beam shape is seen in the lower left inset.}
\label{SDC335_160mu:fig}%
\end{figure}

\begin{table*}
\caption[]{Observed results of the SDC335 milimetre and protostellar cores used in this study. }
\begin{center}
\begin{tabular}{c c c c c c c c c}
\hline
\hline
Instrument & Freq. & $\lambda$ & Resolution$^{\dagger}$ &\multicolumn{4}{c}{$S_{int}$} \\
& [Hz] & [m] & [\asec] &\multicolumn{4}{c}{[Jy]}\\ 
\cline{5-8}
 & & & & MM1 & \mma\ &\mmb\ & MM2  \\
\hline \noalign {\smallskip}
PACS & 4.28 (12) & 70.0 (-6) & 5.0 & 792 $\pm$ 158 & - & - & 604 $\pm$ 121 \\
PACS & 1.87 (12) & 160.0 (-6) & 12.0 & 1456 $\pm$ 582.0 & - & - & 961 $\pm$ 384 \\
LABOCA & 3.45 (11) & 870.0 (-6)  & 18.0 & 8.92 $\pm$ 5.34 & - & - & 1.45 $\pm$ 0.87\\
ALMA & 9.37 (10) & 3.20 (-3)  & 5.6 $\times$ 4.0 & 1.01 $\pm$ 0.10 (-1) & - & - & 1.20 $\pm$ 0.20 (-2) \\
ATCA  & 4.59 (10) & 6.54 (-3) & 3.2 $\times$ 0.8 & 6.85 $\pm$ 1.76 (-3) & & - & $<$0.92 (-4)$^{\star}$ \\
ATCA  & 4.32 (10) & 6.95 (-3)  & 3.5 $\times$ 0.9 & 6.56 $\pm$ 1.37 (-3) & 3.14 $\pm$ 0.91 (-3) & - & $<$0.89 (-4)$^{\star}$ \\
ATCA  & 2.50 (10) & 1.20 (-2)  & 1.9 $\times$ 1.0 & 2.21 $\pm$ 0.32 (-3) & 1.41 $\pm$ 0.17 (-3) & 7.12 $\pm$ 0.12 (-4) & 7.17 $\pm$ 0.99 (-4) \\
ATCA  & 2.30 (10) & 1.30 (-2)  & 2.1 $\times$ 1.1 & 2.48 $\pm$ 0.35 (-3) & 1.76 $\pm$ 0.22 (-3) & 6.02 $\pm$ 0.12 (-4) & 7.24 $\pm$ 1.10 (-4) \\
ATCA  & 8.00 (09) & 3.75 (-2) & 2.3 $\times$ 1.0 & 6.07 $\pm$ 0.98 (-4) & 3.02 $\pm$ 0.39 (-4) & 3.00 $\pm$ 0.36 (-4) & 3.59 $\pm$ 0.42 (-4)\\
ATCA  & 6.00 (09) & 5.00 (-2) & 2.9 $\times$ 1.4 & 3.33 $\pm$ 0.52 (-4) & 1.60 $\pm$ 0.36 (-4) & 2.25 $\pm$ 0.25 (-4) & 2.72 $\pm$ 0.34 (-4)\\
\hline
\end{tabular}
\end{center}
\begin{center}
{\small{$\dagger$, for interferometers resolution given is the synthesised beam major and minor axes.\\ $\star$, 3$\sigma$ upper limits on the none detection of MM2.\\  Columns 2, 3 and 5-8 values in brackets are the order of magnitude in base 10.}}
\end{center}
\label{Results:tab}
\end{table*}

Centimetre wavelength (6 and 8~GHz) observations were made using the ATCA over 13 hours on 9$^{th}/10^{th}$ June 2013. During these observations the ATCA was in the `6C' array configuration, giving a \textit{uv}-range from $\sim$2 to 140k$\lambda$. These observations consisted purely of continuum observations with the CABB system \citep[][]{CABBpaper} set up to observe a 2~GHz bandwidth centred at both 6 and 8~GHz simultaneously. PKS1934-638 was used for flux and bandpass calibration and PKS1646-50 for phase calibration.

Millimetre wavelength (23 and 25~GHz) observations were taken in a single 11 hour session on 13$^{th}$ September 2013, with the ATCA in the `1.5A'  array configuration. During processing, data from the longest ATCA baseline (all antenna 6 baselines) was flagged out to allow better matching of \textit{uv}-range to the 6 and 8~GHz data; the millimetre wavelength data therefore have a {uv}-range from $\sim$9 to 127k$\lambda$. The CABB system was configured to observe two 2~GHz continuum bands centred at 23 and 25~GHz plus seven `zoom' bands to cover six ammonia transitions and the water maser at 22~GHz. Only the continuum are discussed in this paper, the ammonia and water maser data will appear in a later work. The flux and phase calibrators were the same as the 6 and 8~GHz observations with PKS 1253-055 used for bandpass calibration. 

These ATCA data were reduced using the data reduction package \texttt{MIRIAD} with standard ATNF reduction strategies. The task \texttt{MFCLEAN}\footnote{http://www.atnf.csiro.au/computing/software/miriad/doc/mfclean.html} was used to CLEAN the image owing to the large bandwidths of the data. We refer the reader to the \texttt{MIRIAD} user guide\footnote{http://www.atnf.csiro.au/computing/software/miriad/userguide/} for more information. 

The beam and sensitivity characteristics of these observations are given in Table \ref{ObsParams:tab}, including the \citet{Peretto13} ALMA observations of this object for comparison. Integrated flux density measurements from these observations were obtained by fitting Gaussians to each source using the MIRIAD \texttt{IMFIT} task, with the exception of \mma\ at 6~GHz where the slight blending of the two MM1 sources led to unreliable gaussian fits, as such values for this source were measured ``by hand'' using a defined polygonal area in CASA \citep{CASAREF}  (leading to the increased uncertainty in the 6~GHz integrated flux density value). The integrated flux density values are shown in Table \ref{Results:tab}.

\subsection{Additional data}
\label{ancilldata:sec}
In addition to the ATCA observations listed above, data for SDC335 was available at 7mm from ATCA (Avison, Cunningham \& Fuller, \textit{in prep.} Note only MM1 is detected in these data) and ALMA at 3.2mm \citep{Peretto13}. Archival data from \textit{Spitzer}\footnote{Data from Spitzer is shown within Figures \ref{MM1SED:fig} and \ref{MM2SED:fig} for illustrative purposes.}(MIPS and GLIMPSE), MSX, and APEX (LABOCA) have also been used within this work. 

We also made use of \textit{Herschel} data from the "Herschel infrared Galactic Plane Survey" (Hi-GAL, Molinari et al. 2010), which has mapped the whole Galactic Plane covering the wavelength range $70\leq\lambda\leq 500\ \mu$m using both the PACS \citep{Poglitsch10} and SPIRE \citep{Griffin10} photometry instruments in parallel mode on board of the \textit{Herschel space observatory} \citep{Pilbratt10}. We limited our analysis to the PACS 70 and 160 $\mu$m data, which have the highest spatial resolution ($\simeq5"$ and $\simeq12"$ respectively).

The source extraction and photometry of the \textit{Herschel} data has been carried out with \textit{Hyper}, an enhanced  aperture photometry code specifically designed to account for high background variability and source crowding of the Galactic Plane data \citep{Traficante14}. The background is evaluated locally for each source assuming different polynomial orders (from zero to the fifth order) to model the background fluctuations. The polynomial background-subtracted map with the lowest \textit{rms} is taken as reference to evaluate the source flux, and the equivalent polynomial background order as the best fit to describe the background fluctuations in the Hi-GAL maps.

In the case of overlapping sources, as is the case of MM1 and MM2, prior to evaluating each source flux in the aperture region the algorithm models all the sources with a multi-Gaussian fit and subtracts the model of the companions. Within SDC335 the proximity of the brighter MM1 core to MM2 at the resolution of \textit{Herschel} means that the two cores are blended in both the 1.9 and 4.3THz (160 and 70\mewm) \textit{Herschel} maps (see Figure \ref{SDC335_160mu:fig}). It is the data points from these maps which most constrain the SED fit (e.g.  Figure \ref{MM2SED:fig}) discussed later in this work. The flux estimation for each core can be affected by the flux of the overlapping companion or by a coarse background estimation. Although \textit{Hyper} can estimate fluxes with a precision of few percent even in complex background regions and in the presence of heavily blended sources \citep{Traficante14}, we decided to be conservative and assume a flux uncertainty of $\pm$20\% of the photometric \textit{Herschel} flux values.

The source fluxes extracted with \textit{Hyper} have been corrected for aperture corrections. The integrated fluxes from this extraction, $S_{int}$, are shown in Table \ref{Results:tab}.

\begin{figure*}
\centering
\includegraphics[scale=0.55]{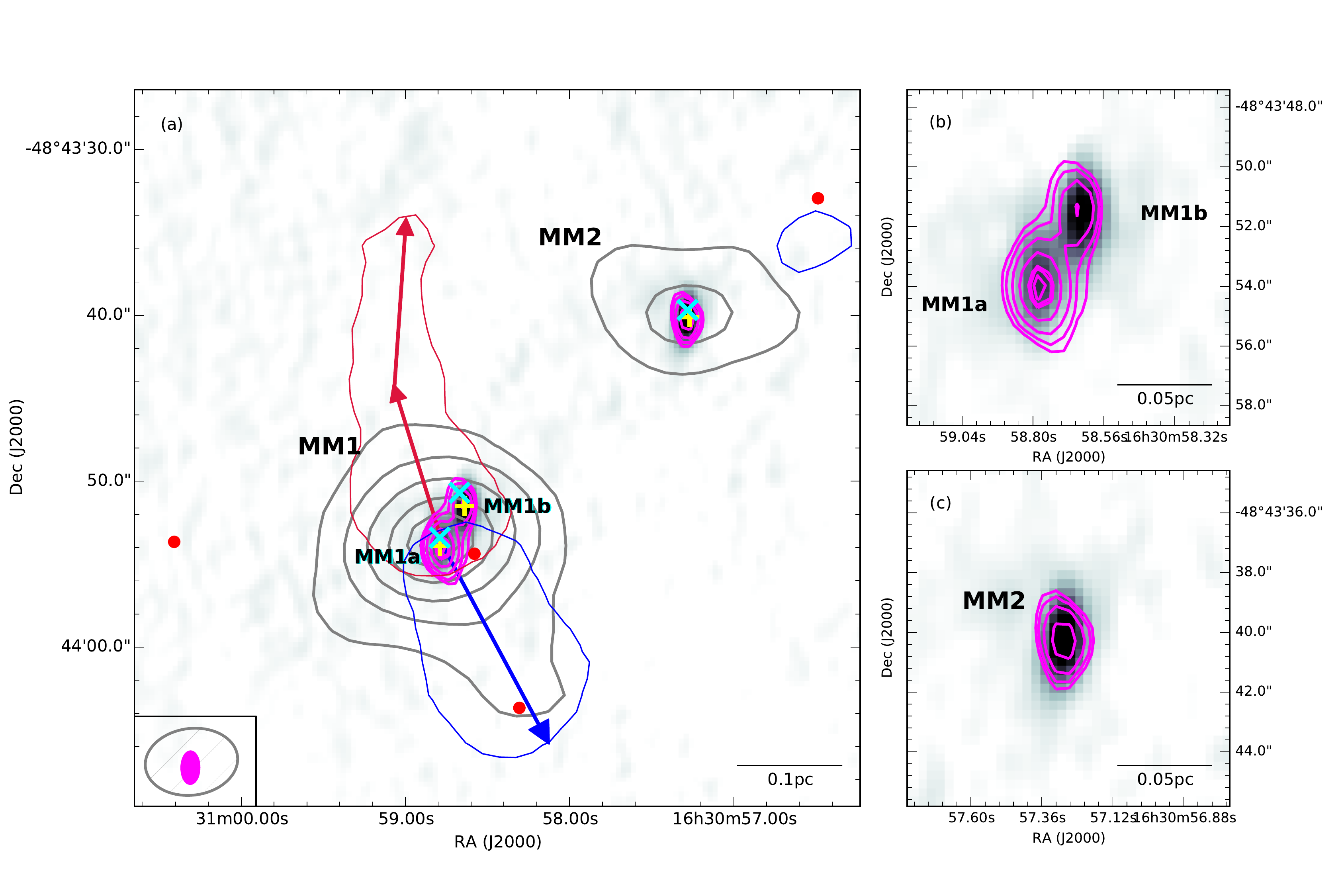}
\caption{\textit{(a)} The SDC335 IRDC as viewed in the radio/mm continuum. Colour-scale the 8~GHz (ATCA) overlaid with ATCA 25~GHz (magenta contours at 15, 20, 30, 50, 75, 80 and 90\% of the peak flux) and the ALMA Cycle 0 3.2mm continuum (grey contours at 2.5, 10, 25, 50, 75 and 90\% of the peak flux).  The location of maser emission in SDC335 are indicated by \textit{$\times$}, {$\bullet$} and \textbf{$+$}  for Class~II (6.7~GHz) and Class~I (44~GHz) \meth\  and \water\ (22~GHz) masers, respectively. The red/blue contours and red/blue arrows represent the 5\% of peak HNC emission boundary and direction of the HNC molecular outflow (Pineda et al., \textit{in prep}) see main text for more detail). The synthesised beam shape for both the ALMA data (grey hatched) and ATCA 25~GHz (solid magenta) are shown in the lower left. \textit{(b) \& (c)} Zoom in of the central regions of MM1 and MM2 respectively, showing 8 and 25~GHz continuum as per main image.}
\label{SDC335pretty:fig}%
\end{figure*}

\section{Results and analysis}\vspace{0.5cm}
SDC335 as viewed in the radio/mm of our observations is seen in Figure \ref{SDC335pretty:fig}. Comparing our ATCA observations with the ALMA data we observe sources at the position of the two mm-cores, MM1 and MM2, in SDC335. However, given the increased resolution of our ATCA data we find that the MM1 core comprises two sources separated by  $\sim$3\asec. We name these \mma\ and \mmb\ , with \mma\ being the more southerly of the pair (see Figure \ref{SDC335pretty:fig} b). The two fragments are situated 2.7\asec\ from one another, a physical separation of $\sim$9000 AU at the distance of SDC335. MM2, the less massive mm-core in SDC335 remains a single source at this increased resolution. The position of each radio core is given in Table \ref{corepos:tab}.  

The red/blue contours and arrows in Figure 3 represent the 5\% of peak emission boundary and direction of the HNC molecular outflow discovered in the ALMA Cycle 0 data (to be discussed in full in Pineda et al. \textit{in prep}), with the colours denoting red and blue shifted emission respectively. Note there is a slight kink in the red shifted outflow, thus the two arrows for this feature. Of interest to the work in this paper is the fact that the position of maximal overlap between the two outflow arms coincides, to within 0.5\asec, with the position of peak continuum flux from \mma. As such we assume that \mma\ is the source which drives the molecular outflow.

In Figure \ref{SDC335pretty:fig} we also show the positions of three maser species: Class I (44~GHz) and Class II (6.7~GHz) \meth\, (with positions from Avison, Cunningham \& Fuller, \textit{in prep.} and \citet{MMB330to345}, respectively) and 22~GHz \water\ (positions from \citet{Breenwater10} and the 22~GHz spectral data from our ATCA observations). 
Class II \meth\ and \water\ masers are typically associated with the central regions of protostellar objects. Class II \meth\ masers are seen to trace discs or disc-like features when observed on milli-arcsecond scales \citep[e.g.][]{Bartkiewicz09} and are uniquely associated with regions of massive star formation. \water\ masers are typically associated with outflow regions in both high and low mass star forming regions \citep{Fish07}, particularly the inner regions ($\sim$6200AU) thereof \citep[][and references therein]{Breenwater10} when the \water\ and class II \meth\ masers are co-spatial. As such, each pairing of these maser species (one for each of the three radio continuum source in SDC335), is indicative of the position of a forming massive star.

Class I \meth\ masers have been found to trace the larger scale molecular outflows of protostellar systems \citep{Kurtz04, Cyganowski09}. Two of the four class I \meth\ maser spots in SDC335 inhabit the blue outflow arm originating from \mma. A third spot is adjacent to the HNC blue lobe observed near MM2.

Given the assumed outflow origin, the angular offset between the outflow direction and a line connecting \mma\ and \mmb, and the presence of both a \water\ and class II \meth\ maser associated with the peak continuum flux of MM1a and b, we consider these two sources independent objects. 

Our analysis focuses on constraining the properties of \mma, \mmb\ and MM2.

\begin{table}
\caption[]{The positions of the radio cores observed in SDC335.}
\begin{center}
\begin{tabular}{ c c c}
\hline
\hline
 Source & \multicolumn{2}{ c}{J2000}   \\
 & Right Ascension &  Declination  \\
 &  [ hh:mm:ss ] &  [ $^{\circ}$ : $^{\prime}$ : \asec] \\
\hline \noalign {\smallskip}
\mma & 16:30:58.765 & -48:43:54.00 \\
\mmb & 16:30:58.638 & -48:43:51.70  \\
MM2 & 16:30:57.291  & -48:43:40.21  \\
\hline
\end{tabular}
\end{center}
\label{corepos:tab}
\end{table}

\subsection{Free-free component}

\begin{figure*}
\centering
\includegraphics[scale=0.55]{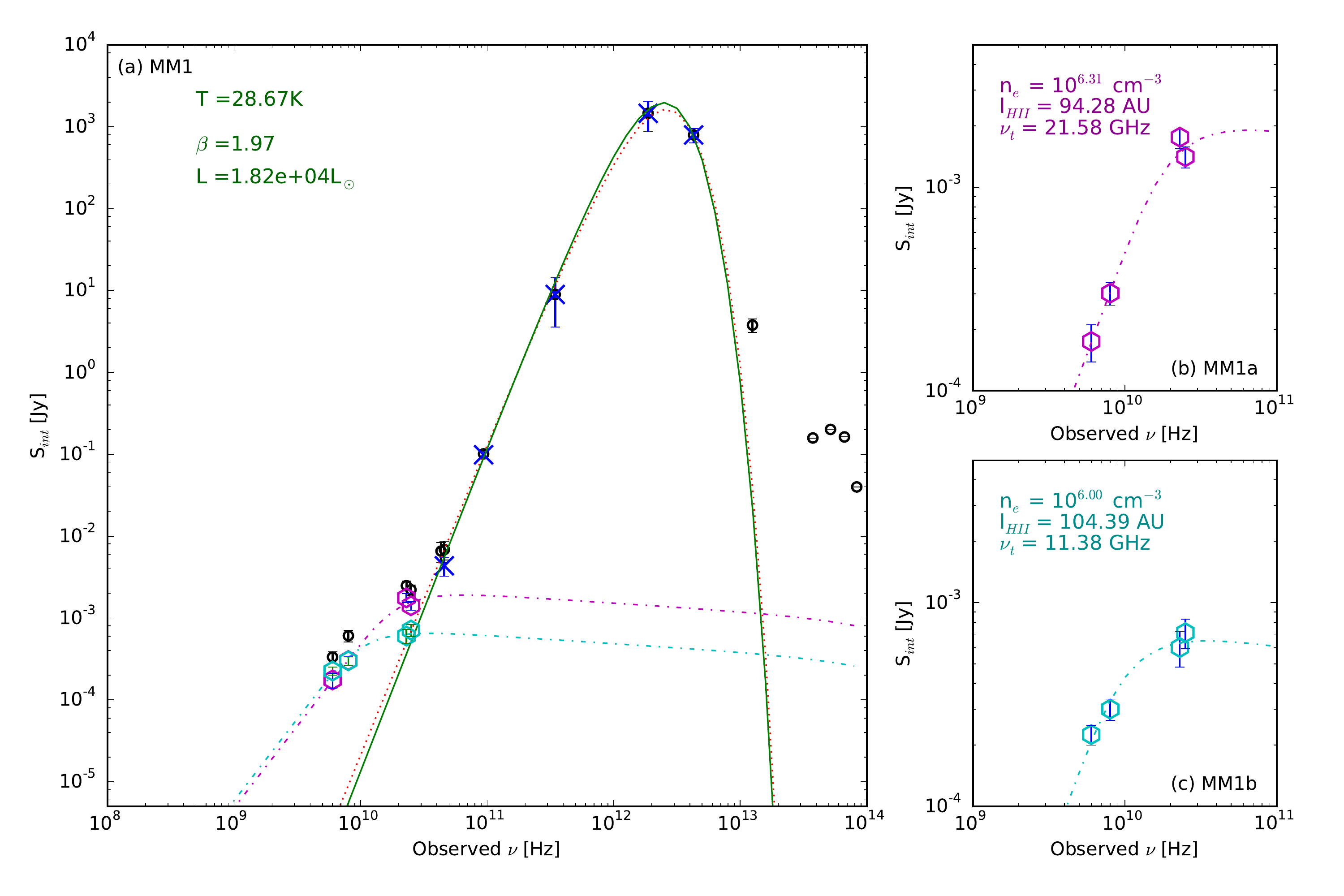}
\caption{ Spectral energy distribution of the MM1 core on $\sim$5\asec scales. $a)$ The black $\circ$ give the fluxes at each frequency, the blue $\times$ are the data used for the blackbody fit, note that at 23 and 25~GHz the free-free component has be subtracted. The black-body fit with and without the free-free component subtraction are given by the solid green and red dotted lines respectively. The magenta and cyan dot-dash lines and hexagons give the radio fluxes and follows the free-free fits for \mma\ and \mmb\ , respectively. The free-free emission data for \mma\ and \mmb\ can be seen separated in $b)$ and $c)$.}
\label{MM1SED:fig}%
\end{figure*}

\begin{figure}
\centering
\includegraphics[scale=0.5]{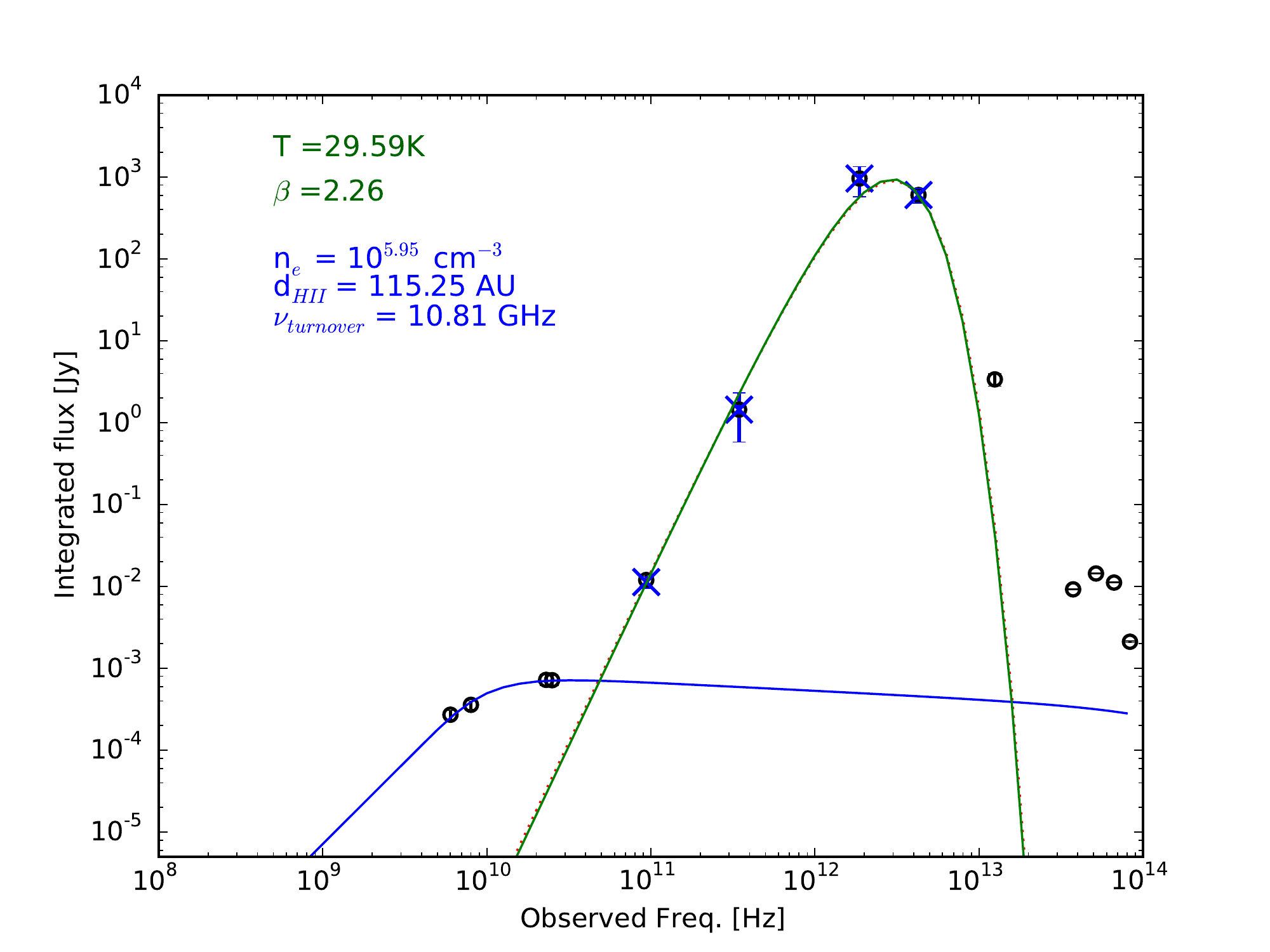}
\caption{Spectral energy distribution of the MM2 core. Symbols as Figure \ref{MM1SED:fig}$a$.}
\label{MM2SED:fig}%
\end{figure}

The three compact \hii\ regions detected in our ATCA observations are observed with sufficient resolution to create partial SEDs from 6 to 25~GHz, these can be seen in Figure \ref{MM1SED:fig} (\textit{b}) and (\textit{c}) and Figure \ref{MM2SED:fig}. We calculate for each sources spectral index, $\alpha$ (where $ S_{int} \propto \nu^\alpha$), with $S_{int}$ being the integrated flux of each source and $\nu$ frequency, the $\alpha$ values are given in Table \ref{AlphaProperties:tab}.  One can consider $\alpha$ as an indicator of the free-free emission turnover frequency.  The turnover frequency of free-free emission being the frequency at which a source transitions from optically thick to optically thin, i.e. where the source optical depth equals 1. Below this frequency a free-free emission region is optically thick and $\alpha = 2$, toward the turnover frequency $\alpha$ reduces and for optically thin sources $\alpha$ approaches $-0.1$ \citep{Kurtz05}.

Each source is optically thick between 6 and 8~GHz. \mma\ appears to remain at least partially optically thick between 8 and 23~GHz and the lower $\alpha$ values between these frequencies for \mmb\ and MM2 (compared to \mma) are also suggestive of partial optical thickness. However, these lower $\alpha$ values do allow for the turnover frequency to occur between these 8 and 23~GHz for both \mmb\ and MM2. 

MM2 is the only source for which there is insignificant contamination from thermal dust emission at 25~GHz, (contrast Figure \ref{MM1SED:fig} and Figure \ref{MM2SED:fig}), the $\alpha$ value between 23 and 25~GHz shows this \hii\ region has become optically thin at these frequencies.

Based upon these spectral indices we consider the nature of the free-free emission, specifically considering the ambiguity between compact \hii\ regions and collimated ionized jets at these resolutions. The steep spectral index of MM1a between 6-8 and 8-23~GHz is typical of that of a compact \hii\  region \citep{Kurtz05} and steeper than the expected value for collimated jet emission ($\alpha$=0.25-1.1, \citet{Reynolds86}). Thus we classify \mma\ as a hyper compact \hii\ (HC\hii) region.

The spectral indices of \mmb\ and MM2 of $\sim$1.0 and $\sim$0.6 at 6-8 and 8-23~GHz respectively in both sources, fall within the characteristic range of both collimated jet emission \citep{Reynolds86} and photo-ionized compact \hii\ regions in the transition from optically thick to thin. In the case of \mmb, given its misalignment to the molecular outflow originating from \mma\ and centrally located (potentially disc tracing) \meth\ maser we assume \mmb\ is not part of a radio jet within the MM1 mm-core and instead favour the interpretation that this source is also compact \hii\ region ionized by a separate source to \mma. We also assume the compact \hii\ region interpretation for MM2, based on the lack of a firm detection of a nearby strong collimated outflow. However, within the scope of these data either the radio jet or compact \hii\ region interpretation may be valid for both \mmb\ and MM2. The ultimate result of this ambiguity is that the stellar mass limits calculated assuming a compact \hii\ region, ergo a photoionization origin for the radio emission, will not hold true for radio jets. However the association of \mmb\ and MM2 with class II \meth\ masers confirms that they are forming massive stars.

\begin{table}
\caption[]{Spectral index properties of the cores within SDC335. We do not include the $\alpha$ value between 23 and 25~GHz for sources \mma\ and \mmb, as it is possible the value is contaminated by some thermal emission (see Figure \ref{MM1SED:fig} (\textit{a})). }
\begin{center}
\begin{tabular}{@{}c c c c }
\hline
\hline
$\nu$  & \mma\ &  \mmb\ & MM2  \\
interval & $\alpha$ & $\alpha$ & $\alpha$  \\
\hline
6 - 8 GHz & 1.90$\pm$0.40 & 1.01$\pm$0.12 & 0.96$\pm$0.11  \\
8 - 23 GHz & 1.67$\pm$0.20 & 0.66$\pm$0.13 & 0.67$\pm$0.10 \\
23 - 25 GHz & - & - & -0.11$\pm$0.02 \\
\hline
\end{tabular}
\end{center}
 \label{AlphaProperties:tab}
 \end{table}

\subsubsection{Free-free fitting technique}
\label{fffit:sec}
To further this analysis we fit the free-free component using the 6, 8, 23 and 25~GHz ATCA data, implementing a non-linear least square regression algorithm. The fitted parameters are the free-free turnover frequency, $\nu_t$, (where the optical depth, $\tau=1$) and the physical diameter of the source, $l$. Setting $\tau=1$ and fixing $T_e$ to the fiducial 10,000K \cite[e.g][]{Spitzer78,SanchezMonge11}, emission measure, ($EM$), values for a given turnover frequency were calculated using the approximate formula for the optical depth of free-free emission \citep[][and references therein]{MezgerHenderson67}:

\begin{equation}
\centering
\tau_{ff} \approx 0.08235\ T_e^{-1.35}\  \Bigg( \frac{\nu}{\rm{GHz}} \Bigg)^{-2.1} \Bigg( \frac{EM}{\rm{pc~ cm}^{-6}} \Bigg)\rm{~}.
\label{HIIoptdepth:eqn}
\end{equation}

\noindent Based on this $EM$ value the optical depth, $\tau_{ff}$, following Equation \ref{HIIoptdepth:eqn}, was used in Equation \ref{fitflux:eqn},
\begin{equation}
\centering
S_{ff-fit}=\Omega_{ff} B_{\nu}(T_e)(1-e^{-\tau_{ff}})\rm{~},
\label{fitflux:eqn}
\end{equation}

\noindent to fit the flux density of the data.  Here $\Omega_{ff}$ is the solid angle subtended by the source and $B_{\nu}(T_{e})$ the Planck distribution value at a given frequency at $T_e$. $S_{ff-fit}$ was then fit to the observed fluxes with the best fit giving values for $\nu_t$, $EM$ and $l$.  Note that the $EM$ value calculated at the turnover frequency is independent of any knowledge of the sources physical dimensions meaning there is no degeneracy in the value of $l$ which is used only in calculating the value of $\Omega_{ff}$.

\subsubsection{Fitted Properties}
\label{fffitprop:sec}
The fitted properites for the three \hii\ regions seen in SDC335 are given in Table \ref{CoreProperties:tab}. Each object has fitted size of < 0.5mpc ($\simeq$110AU), a value much smaller than that of the upper size limit associated with the hyper-compact \hii\ region classification ($\lesssim$30mpc e.g \citealt{Kurtz05}). Sources of similar size have been observed in high mass star forming regions, e.g. W3 IRS 5 \citep{Wilson03} with sources d1/d2 and f  both measuring deconvolved radii of 0.8mpc ($\simeq$ 165AU) using the VLA at 1.3 and 0.7cm. \citet{Zhu13} observed the HC\hii\ region NGC7538 IRS1 with the SMA and CARMA (at 1.3 and 3.4mm) and resolved it into a compact component <270AU ($\simeq$1.3mpc) and an extended tail of 2000AU ($\simeq$ 9mpc). Also, \citet{SanchezMonge11} in their VLA search for new HC\hii\ regions (in a methanol maser selected sample) found six new HC\hii\ candidates with derived sizes of between 1 and 10mpc ($\simeq$206-2060 AU).

The fitted optically thick/thin turn over frequency for the three SDC335 \hii\ regions are in agreement with the values expected from the spectral indices calculated between each of our observing frequencies. \mma\ is the densest of the 3 \hii\ regions in SDC335 remaining optically thick to frequencies $\geq$20~GHz. \mmb\ and MM2 exhibit similar properties to one another with fitted turnovers of $\sim$ 11~GHz. From the fitted values for each of our compact radio sources it seems likely that SDC335 harbours three HC\hii\ regions. The fit to \mma\ is the least robust as the data for this source allows a rise in flux beyond 25~GHz. However, the fitting technique limits the turnover frequency to be within the frequency range of the data. This issue is addressed further in \S \ref{postprior:sec}.

\subsubsection{Derived properties}
\label{Derivprop:sec}

For each source we calculated an electron number density, $n_e$,  and the number density of Lyman $\alpha$ photons being emitted. The latter allowing the association of the three HC\hii\ regions with a zero-age main-sequence (ZAMS) stellar spectral type. Assuming a cylindrical homogeneous \hii\ region $EM$ is related to electron number density by $EM$ =$\int_{0}^{l} n_{e}^2 dl \approx n_{e}^2l$. Calculated values for $n_e$ are found in Table \ref{CoreProperties:tab} and again exhibit values associated with HC\hii\ ($\gtrsim 10^6$cm$^{-3}$).

The Lyman continuum flux assuming optically thin emission is given by: 
\vspace{0.5cm}
\begin{equation}
\centering
\begin{split}
N_{Lym}=7.6\times10^{46}\ \alpha^{-1}(\nu,T_e)  \ \Bigg(\frac{T_e}{10^4\rm{K}}\Bigg)^{0.35}
\\ \times \  \Bigg(\frac{\nu}{\rm{GHz}}\Bigg)^{0.1} \Bigg(\frac{S_{int, cont}}{\rm{Jy}}\Bigg)\Bigg(\frac{D}{\rm{kpc}}\Bigg)^{2.0} \rm{ photons~s}^{-1}
\end{split}
\label{Nlym:eqn}
\end{equation}

\noindent \citep{Zhu13}, where $S_{int, cont}$ is the radio continuum flux density at a given frequency and $D$ the distance to SDC335 (i.e. 3.25kpc). For each source we calculate $N_{Lym}$ using the 23~GHz continuum integrated flux density value. It is important to note that as \mma\ is optically thick beyond 25~GHz, use of the 23 (or indeed 25)~GHz  will lead to an underestimate in $N_{Lym}$ for this source meaning our calculated value is a lower limit. As previously, we set $T_e$=10,000K, giving $\alpha(\nu,T_e)=0.966$ \citep{MezgerHenderson67}.
 
The $N_{Lym}$ values were then compared with the Lyman Continuum values for ZAMS stars given in \citet{Davies11} and spectral types from \citep{Mottram11}. From these values the protostellar cores in SDC335 appear to have \textit{currently} formed a single B1.5 ($M_*$ = 9.0 \solmass) type star in the MM2 core, while the MM1 core has formed at least two stars one of B1.5 (\mmb) spectral type and a second of \textit{at least} spectral type B1-B1.5 ($M_*$ =10.0 \solmass) for \mma.  These spectral classifications are based on the properties of the protostars as they are observed now were they at the ZAMS stage and not still actively accreting. 

\begin{table*}
\caption[]{Properties of the cores within SDC335 from the SED fitting.}
\begin{center}
\begin{tabular}{@{}l  c c c c c c c c c}
\hline
\hline
Object & $EM$ & $n_e$ & $l_{\hii}$  & $N_{Lym}^{\dagger}$ &  $L _{bol}$ & $T_{d}$ & $\beta$ & $M_{gas}$\\
 & [pc cm$^{-6}$] & [cm$^{-3}$] & [AU] & [photon s$^{-1}$] & [L$_{\odot}$]  & [K] & & [\solmass]\\
\hline
MM1 & -  & -  & -  & - & 1.8$\times10^{4}$ & 28.7($\pm$2.3) & 1.97($\pm$0.15) & 763($^{+165}_{-171}$) \\
\textit{MM1$_{p}$} & -  & -  & -  & - & \textit{1.6$\times10^{4}$} & \textit{30.0($\pm$1.7)} & \textit{1.75($\pm$0.11)} & \textit{429($^{+104}_{-109}$)} \\
\mma & 1.9$(\pm0.5)\times10^9$  & 2.1$(\pm0.3)\times10^6$  & 94.28($\pm$5.7)   & >2.0$(^{+0.5}_{-0.5})\times10^{45}$ & -  & - & - & -  \\
\textit{MM1a$_{p}$} & \textit{2.0$(\pm$0.4)$\times10^{9}$} & \textit{2.1$(\pm$0.2)$\times10^6$}& \textit{93.96($\pm$7.1)} & \textit{''} & -  & - & - & - \\
\mmb & 5.0$(\pm1.0)\times10^8$ & 1.0$(\pm0.1)\times10^6$ & 104.4($\pm$5.2)  & 6.8$(^{+1.9}_{-2.0})\times10^{44}$ & - & - & - & - \\
MM2 & 4.5$(\pm0.5)\times10^8$ &8.9$(\pm0.5)\times10^5$ & 115.2($\pm$4.7)  & 8.2$(^{+1.9}_{-2.0})\times10^{44}$& 9.9$\times10^{3}$ & 29.6($\pm$7.2) & 2.26($\pm$0.52) & 173($^{+65}_{-66}$) \\
\hline
\end{tabular}
\end{center}
\begin{center}
{\small{$\dagger$:  $N_{Lym}$ values are lower limits, see text.\\ The values shown in rows \textit{MM1$_p$} and \textit{MM1a$_{p}$} are those which include the 43~GHz data point in the free-free emission SED fitting, see \S \ref{postprior:sec} for more details.}}
\end{center}
\label{CoreProperties:tab}
\end{table*}

\subsection{Spectral Energy Distribution of SDC335 protostellar cores}

Combining the multi-frequency data available (from this work and ancillary data described in \S \ref{ancilldata:sec}), and listed in Table \ref{Results:tab}, we create spectral energy distributions (SEDs) for the dust emission from MM1 and MM2.  These were created by first subtracting the best fit free-free emission from the thermal emission at all frequencies used in the fit (25~GHz to 4.3~THz (70\mewm) for each source) and then fitting the SED to these corrected data.

For the purpose of SED fitting MM1 is treated as 2 objects (\mma\ and \mmb) at frequencies between 6 and 25~GHz and as a single object above these frequencies where the resolution is 5\asec\ or greater. As the respective resolutions of the data used in this fitting vary as a function of frequency, we conservatively compensate for potential inclusion of excess flux in lower resolution data. To achieve this the uncertainties of the data points (presented in Table \ref{Results:tab}) with beams greater than 6\asec\ have been scaled assuming a density profile of $r^{-2}$ for these sources \citep[see][Table 1.]{Peretto13}.

The modified blackbody fitted is of the form:

\begin{equation}
\centering
S_{bb-fit}=\Omega_c B_{\nu}(T_d)(1- e^{-\tau_{bb}})
\label{bbfitflux:eqn}
\end{equation}

\noindent where $B_{\nu}(T_{d})$ is the Planck distribution value at a given frequency, $\nu$, and a dust temperature $T_d$. $\tau_{bb}$ is the optical depth of dust and has the form $\left(\frac{\nu}{\nu_0}\right)^{\beta}$, with $\nu_0$ as the frequency at which the optical depth is unity. Finally, $\Omega_c$ is the solid angle subtended by the dust dominated region, for the purposes of fitting we fix $\Omega_c$ to the area subtended from the \textit{Herschel} data using the \textit{Hyper} photometry code \citep{Traficante14} which has radius 11.4\asec.  

The free parameters in our fitting routine are, $T_d$, $\nu_0$ and $\beta$, and are given priors of 25.0K, 4.3 THz and 2.0 respectively. The resulting best fit parameters are then used in a direct integration of the SED between 100 kHz and 100 THz (3km to 3\mewm) to calculate the source bolometric luminosity. We also refine the mass of the mm-cores MM1 and MM2 using the 3.2mm ALMA flux densities (see Table \ref{Results:tab}), at deconvolved size scales of 0.054pc and 0.057pc (following \citealt{Peretto13}), applying our fitted $T_d$, $\nu_0$ and $\beta$ in the mass flux relation,

\begin{equation}
\centering
M_{gas}=\frac{D^2S_{3.2}}{\kappa_{3.2}B_{3.2}(T_d)}
\label{massflux:eqn}
\end{equation}

\noindent and using the dust opacity law $\kappa_{\lambda}=0.1 \times \left( \frac{\lambda}{0.3mm}\right)^{-\beta} $cm$^{2}$ g$^{-1}$, assuming a dust to gas ratio of 0.01 \citep{Hildebrand83,Beckwith90}.

\subsubsection{SED fitting results}
\label{SEDcores:sec}
Table \ref{CoreProperties:tab} presents the parameters derived from our SED fitting. Overall the values fitted and derived values from our SEDs are typical of those of a massive protostellar core in early stages of star formation \citep[e.g][]{Rathborne08, Beuther10,Rathborne10}. 

\subsubsection{Posteriori fitting of MM1a}
\label{postprior:sec}
As noted in \S \ref{fffit:sec} the free-free emission from \mma\ is the least well constrained with the turnover potentially occurring above the 25~GHz data point used in the inital fit. To attempt to overcome this we use the SED fitted results for MM1 (T=28.7K and $\beta$=1.97) following the above method. We then adjust the 43~GHz data point and refit the free-free emission from \mma\ including this data point. The model fit of \mmb\ is kept as constant within these fits. The fitting of free-free emission data and then the full (free-free subtracted) SED proceeds iteratively until best fit values for \mma\ in free-free ($l$, $\nu_t$ and EM) and new best fit values for MM1 as a single object (T and $\beta$) are found. These new values can be found in Table \ref{CoreProperties:tab} labelled subscript `p' and represented in italics.

The new fitted properties for MM1 and \mma\ are consistent with the initial fitted values for each source, which may suggest a robust fit to these data or highlight the detrimental effect of the larger error associated with the 43~GHz data point compared to those from the primary data (6-25~GHz ATCA) for this work. As such we consider only the initial fit values throughout the remainder of this work.

\section{Discussion}
\subsection{The evolutionary status of the ionising sources in SDC335}

In $\S$\ref{Derivprop:sec} we calculate a Lyman photon flux for each HC\hii\ and from this associate a ZAMS spectral type to the source ionising each HC\hii. We now consider the evolutionary status of the three ionising sources in SDC335 with respect to the calculated $L_{bol}$ values for the MM1 and MM2 cores ($\S$\ref{SEDcores:sec}) and the protostellar evolutionary tracers associated with each source compared to the characteristics that the ZAMS spectral type would apply. 

A star arrives at the Zero Age Main Sequence once hydrogen burning has commenced. Within SDC335 we expect accretion to be ongoing, as evidenced by the presence molecular outflow from the MM1 mm-core (Pineda et al. \textit{in prep.}), and each ionising source exhibiting a 22~GHz water maser (\citet{Breenwater10}; Avison et al. \textit{in prep.}), a maser species associated with outflows from both low and high mass YSOs \citep{Fish07}. In addition to these, an Extended Green Object (EGO, \citealt{Cyganowski08}), another outflow tracer, is also present at the centre of SDC335 filamentary structure. The global collapse of SDC335 also supports the ongoing accretion of the protostars ionising the three detected HC\hii.

Accretion rates of between $\geq \times10^{-4}$ and up to $few \times10^{-3}$ \solmass\ yr$^{-1}$  \citep[e.g.][]{Fuller05, Churchwell10,Klaassen12, Rygl13, Ana13} have been calculated for massive star precursors (e.g. HMPOs, HC\hii), a greater value than that associated with their lower mass counterparts ($\dot{M_*}\sim \frac{c_s^3}{G}\simeq5\times10^{-6}$ \solmass\ yr$^{-1}$ \citep{Shu77,McKeeTan03turb}). Given such observed values we make the assumption that the infall rate of SDC335 (2.5 $\pm$ 1.0)$\times10^{-3}$ \solmass yr$^{-1}$ \citep{Peretto13} continues unimpeded onto the protostars. During the accretion phase of a young stellar object the total output luminosity is comprised of an accretion component and the intrinsic luminosity, $L_{tot} = L_* + L_{acc}$ with the accretion component taking the form $L_{acc} = \frac{GM_{*}\dot{M}_{*}}{R_*}$.  If we assume the protostars have the ZAMS radii, $R_*$ and luminosity, $L_*$ interpolated from \citet[Table 1]{Davies11} to our calculated $N_{Lym}$ values and use the ZAMS mass-luminosity relation of $L_* \propto M_{*}^{3.5}$ we can calculate a mass for each protostar. With these values we then calculate $L_{tot}$ for MM1 and MM2 (in the case of MM1 the value is the addition of calculated values for \mma\ and \mmb). In both cases $L_{tot}$ exceeds the calculated $L_{bol}$ by a factor $\sim$20, with $L_{acc}$ the dominant luminosity (see Table \ref{ZAMSvalues:tab}). 

\begin{table}
\caption[]{Calculated Luminosity values for \mma, \mmb\ and MM2 using the assumed ZAMS properties of the ionising source each.}
\begin{center}
\begin{tabular}{c c c c}
\hline
\hline
 & $L_*$ & $ L_{acc}$ & $L_{tot, ZAMS}$  \\
Source &  [$L_{\odot}$] &  [$L_{\odot}$] & [$L_{\odot}$] \\
\hline
\mma\ & 5.48$\times10^3$ & 2.22$\times10^5$ & 2.27$\times10^5$\\
\mmb\ & 4.10$\times10^3$ & 2.16$\times10^5$ & 2.20$\times10^5$\\
MM2 & 4.35$\times10^3$ & 2.17$\times10^5$ & 2.21$\times10^5$\\
\hline
\end{tabular}
\end{center}
\label{ZAMSvalues:tab}
\end{table}

Initially we note that the ZAMS $L_*$, without the additional luminosity contribution from $L_{acc}$, for both MM1 and MM2 are within a factor $\sim$ 2 of the measured $L_{bol}$, not a large discrepancy given the uncertainites in the data. However, the number of outflow indicators associated with the protostars (as noted above) excludes the scenario in which the SDC335 protostars have ceased accretion.  Instead we consider two mechanisms by which the discrepancy between $L_{tot,ZAMS}$ and $L_{bol}$ can be addressed.

First, we consider the models for massive protostellar evolution with high accretion rates ($\geq$10$^{-3}$\solmass\ yr$^{-1}$) presented by \citet{HosokawaOmukai09} and \citet{HosokawaYorkeOmukai10} for spherical collapse and accretion via a disc respectively. Under both scenarios massive protostars transition through a number of stages before joining the ZAMS. Notably swelling at masses 6 $\lesssim M_* \lesssim$ 10 \solmass\ and contracting again between 10 $\lesssim M_* \lesssim$ 30 \solmass, (for a constant $\dot{M}\sim 10^{-3}$\solmass\ yr$^{-1}$). During the swelling phase the protostellar radii swells rapidly from a few $R_{\odot}$ to $\sim$100 $R_{\odot}$ whilst the intrinsic luminosity also increases. As a result of the swelling the effective temperature of the protostar is lower than that of a ZAMS star with the same luminosity, with temperatures low enough that there are insufficient photons with energies greater than 13.6eV to create a \hii\ region. 

Again following the $\dot{M}\sim 10^{-3}$\solmass\ yr$^{-1}$ case reported on for both accretion scenarios in \citet{HosokawaOmukai09} and  \citet{HosokawaYorkeOmukai10}, after the swelling of the protostar a period of contraction commences whilst accretion continues. $L_*$ becomes dominant component of $L_{tot}$ over $L_{acc}$ and $R_*$ drops from $\sim$100 $R_{\odot}$ to of the order 10 $R_{\odot}$ as the protostar increases in mass. It is during this phase that the protostellar effective temperature becomes high enough that the flux of ionising photons is sufficient to start the formation of a \hii\ region.

Under this high accretion rate scenario and given that our protostars are displaying ionization but have $L_{bol}<L_{tot, ZAMS}$ we suggest the possibility that the protostars within SDC335 are in the contraction phase of their formation, putting their current masses each at $\geq$ 10\solmass\ (see e.g. \citet[Figure 18]{HosokawaOmukai09}) with their eventual final ZAMS mass higher than this. It is interesting to note that our assumed mass accretion rate (2.5 $\pm$ 1.0)$\times10^{-3}$ \solmass\ yr$^{-1}$ approaches the limiting case of the \citet{HosokawaYorkeOmukai10} model at $\sim$4$\times10^{-3}$ \solmass\ yr$^{-1}$, an accretion rate value at which a second stage of rapid radial expansion occurs halting contraction and steady accretion ceases. In this case the temperature again decreases owing to the radial expansion and no \hii\ regions are created. This provides a maximum limit on the accertion rates of these sources of $\lesssim 4\times10^{-3}$ \solmass\ yr$^{-1}$ \citep{HosokawaYorkeOmukai10}.   

The second mechanism which can address the luminosity discrepancy is that there is a decreased, decreasing or periodic accretion when compared to the $\dot{M_*}$ used in our $L_{acc}$ calculation. The value $\dot{M_*}=2.5\times10^{-3}$\solmass\ yr$^{-3}$ used was found by \citet{Peretto13} and assumes spherical accretion onto the central mm-cores of SDC335 due to the global cloud collapse. It is possible that the accretion rate onto the protostars is lower than into the cores, meaning that the cores are growing at a greater rate than the protostars thus $L_{acc}$ is lower. Indeed, reducing the accretion rate onto the protostars by the factor 20 notes gives an $\dot{M_*}$ of $1.25\times10^{-4}$\solmass\ yr$^{-3}$ which is not unreasonable for a forming massive star.

Decreasing or periodic accretion onto the protostars would lead to fluctuations in $L_{acc}$ over time. These are the most difficult scenarios to address in any quantitative manner as it requires both knowledge of the accretion on to the sources in a more direct way than our current data allow and preferably observation of this accretion or suitable proxies at more than one epoch. Opposing these accretion rate scenarios \citet{Davies11} found in their population synthesis based on the RMS data \citep{Hoare04}  that declining (or indeed constant) accretion rates as a function of time do not well fit the observed massive star population of the galaxy. There is currently no concensus on periodic accretion onto massive protostars within the literature.

Our current data on SDC335 do not allow us to select the likely candidate from these scenarios as being the mechanism responsible for the lower than expected $L_{bol}$. Further observation of the SDC335 protostars targetted at constraining the accretion on smaller scales is required to resolve this issue.

\subsection{Cluster formation in SDC335}

Given that we have discovered three massive protostars within SDC335 it is possible to put limits on the eventual number of stars the IRDC is capable of creating. We employ the power law form of the IMF, $N \propto M^{-\alpha}$, with $\alpha$=2.35 \citep{Salpeter55} and $\alpha$=2.3 \citep{Kroupa02} in the standard mass range of 1-120\solmass. With the criteria that 3 objects have masses above 9.7\solmass, we find that SDC335 has the potential to form a minimum of $\sim$54-60 stars (between 1 and 120\solmass).

Continuing the \citet{Kroupa02} IMF down to a mass of 0.08\solmass\ SDC335 has the potential to form $\sim$1400 stars, with the vast majority of these ($\sim$96\%) in the sub-solar mass regime. Using the average stellar mass from \citet{Kroupa02} Table 2 this would give an approximate total mass in stars, $M_{\star tot}$, of $\sim$285\solmass, a small fraction (5\%) of the total mass of the SDC335 cloud (5500$\pm$800\solmass). Assuming a typical star formation efficiency of $\sim$10 - 30\% for embedded clusters \citep{LadaLada03} one would expect a $M_{\star tot}$ from SDC335 in the range 550-1650\solmass. Keeping in mind the uncertainties on our mass estimates, the discrepancy between this and our IMF calculated value could be the result of our underestimation of the true final masses of the three massive protostellar objects in SDC335. 

\subsubsection{The future of SDC335}
\label{OBClu:lab}

For a young system such as SDC335, in the early stages of its star formation process it is interesting to make comparisons to systems with similar characteristics at different stages of evolution. One such system is the Trapezium Cluster (TC) at the centre of the Orion Nebula Cluster. This extensively studied \textit{relatively} evolved system \citep[see e.g.][]{Hillenbrand97, Muench02}, has some obvious similarities with our younger system. Centered on $\theta^1C$ Ori, TC has a diameter of $\sim$0.3pc \citep{Hillenbrand97} comparable to the separation of the two SDC335 millimetre cores (0.32pc). The eponymous Trapezium stars comprise five OB stars at the heart of the cluster similar to the three protostars exciting the HC\hii\ regions described in this work. Extending out to include the inner 3pc (and encompassing the TC) is the Orion Nebula cluster (ONC), a region of physical size comparable to the whole of SDC335 ($\sim$2.4pc).

Beyond these obvious similarities, the stellar population study of the ONC as reported by \citet{Hillenbrand97} detects $\sim$1600 optically visible stars, with an expected additional 50\% not detectedd due to their embedded nature. Of these sources the authors report on the 60\% of stars for which photometry \textit{and} spectral types are available allowing these objects to be associated with a mass. In the mass range 1-50\solmass\ their sample is complete and contains 137 objects with 6 stars of mass greater than 10\solmass, (the Trapezium stars plus others at a larger distance from $\theta^1C$ Ori). We have shown SDC335 has 3 stars of mass $\gtrsim$ 10\solmass\ and we expect 60 stars in a similar mass range to the 137 stars in ONC, from this it would appear we can expect a similar population in SDC335 to that in the ONC to within a factor $\sim$2. 

It is also interesting to note that in SDC335 all three HC\hii\ regions are found in the IRDCs central region (at the confluence of the filamentary arms). In the more evolved ONC \citet{Hillenbrand97} find evidence for mass segregation with the most massive stars preferentially located within projected radii of $<$0.3pc. In a following paper \citet{HillenbrandHartmann98} find that this distribution of stars is unlikely to be the result of dynamical mass segregation since its timescale is larger than the age of the ONC stars. The authors instead favour a primordial origin of the observed mass segregation. This interpretation is consistent with our SDC335 findings, the three most massive sources lying at the bottom of the gas-dominated, gravitational potential well.

A second comparable region is the NGC6334 I(N) protocluster which has been well studied at mm wavelengths, most recently by \citet{Hunter14}. NGC 6334 I(N) (hereafter simply I(N)) is an IRDC at a distance of 1.3kpc and displays seven compact millimetre continuum sources within a projected diameter of 0.1pc at 2\asec resolution. Further to this, in their study \citet{Hunter14} find 25 likely protostellar objects associated within a physical radius of 0.21pc at 1.3mm. 

There are four radio (5~GHz) continuum sources within the same 0.21pc radius area, three of which are labelled as protostellar candidates and two are directly associated with 1.3mm cores. The limited number of radio continuum detections in I(N) has an interesting corollary with SDC335, scaling the fluxes of the I(N) radio continuum detections to the distance of SDC335 we find that two of the four I(N) sources would be observed as 5$\sigma$ or greater in SDC335. Given this and the fact we do observe three radio continuum sources in SDC335, if we assume I(N) and SDC335 have similar protostellar population density profiles one would expect a similar number of compact mm sources to exist in SDC335. However, currently, the comparatively low resolution of the \citet{Peretto13} ALMA data at 3.2mm compared to that with of the 1.3mm observations of I(N) (coupled with the fact that these two mm frequencies will be tracing different regions around the protostellar populations therein) means we are limited in the our current knowledge of the mm-core population of SDC335 to make a quantative comparison. 

Qualitative comparisons of SDC335 to the NGC 6634 I(N) protocluster and the relatively evolved Trapezium cluster indicates that SDC335 is likely to form a similar population of stars. Further investigation of SDC335 with higher resolution at mm wavelength observations will allow the lower mass/non ionising core population to be detected making these comparisons more robust. 


\section{Conclusions}
We have presented the results of radio continuum observations conducted with the ATCA toward the infrared dark cloud SDC335. Within the two massive protostellar cores observed by ALMA in SDC335 we have observed three compact \hii\ regions. Fitting of the free-free emission of these sources indicates that they have characteristics typical of Hyper Compact \hii\ regions. Based on the fitted and derived characteristics of these HC\hii\ regions we find that the three protostars exciting them display characteristics of ZAMS stars of at minimum, spectral type B1.5 for two of the three and B1-1.5 for the third. This can be considered a lower limit of their characteristics as accretion is still ongoing within SDC335.

With the use of ancillary data, free-free component subtracted SEDs were created allowing the calculation of the bolometric luminosities for the MM1 (treated as a single object at frequencies above 25~GHz) and MM2 cores. We find values from our SED fits to each HC\hii\ region and mm-core which are typical of massive protostellar sources.

Our observed luminosities for the two cores are lower than would be expected from the ZAMS classification derived from the ionized flux. We interpret this result as a consequence of one of two mechanisms reducing the contribution of accretion to the total luminosity.  First, the protostars may be at an early stage of accretion, still with an enlarged but contracting stellar radii after the `swollen core' stage for high accretion rate objects ($\dot{M}_* \geq 10^{-4}$\solmass\ yr$^{-1}$ \citealt{HosokawaOmukai09}) indicating that the protostellar objects exciting our HC\hii\ regions will ultimately become more massive that their current ionising flux indicates. Secondly the assumed accretion rate onto the protostars within the mm-cores MM1 and MM2 maybe lower than onto the cores themselves.

Comparisons between SDC335 and both the realtively evolved Trapezium cluster and the evolving NGC 6334 I(N) protocluster have been drawn to highlight the similarities of SDC335 to these clusters and indicate the potential future of the cluster forming in SDC335. 

Use of future high resolution spectral line and dust continuum observations of SDC335 would allow detection of more precise knowledge of accretion rates of each object and detection of the mass fragmentation and indeed segregation. These would provide more details of how SDC335 is forming a stellar cluster similar to the Trapezium Cluster and NGC 6334 I(N).

\begin{acknowledgements}
The Australia Telescope Compact Array is part of the Australia Telescope which is funded by the Commonwealth of Australia for operation as a National Facility managed by CSIRO. The authors would like to thank the ATCA/ATNF staff who provided excellent help during the observations of the data this paper is based on. We also would like to thank the anonymous referee for their helpful comments on this paper. AA is funded by the STFC at the UK ARC Node. ADC acknowledges funding from the European Research Council for the FP7 ERC starting grant project LOCALSTAR. AT is supported by a consolidated STFC grant to JBCA. JEP is supported by the Swiss National Science Foundation, project number CRSII2\_141880.
\end{acknowledgements}

\bibliographystyle{aa}
\bibliography{../../../Bibliography/Bibliography}

\end{document}